\documentstyle[12pt,a4,psfig]{article}
\begin{document}

\baselineskip 24pt

\title{
$e^{+}e^{-}\rightarrow Z^0Z^0\rightarrow b\overline{b}c\overline{c}$
 events as model independent probe of colour reconnection effects}

\author{  Shi-yuan Li$^{1}$, Zong-guo Si$^1$, 
Qu-bing Xie$^{2}$, Qun Wang$^{2}$\\
 $^{1}$ Department of Physics, Shandong University\\
Jinan, Shandong 250100, P.R.China\footnote{e-mail: tpc@sdu.edu.cn}\\ 
$^{2}$Center of Theoretical Physics, 
CCAST(World Lab)\\
Beijing 100080, P.R.China,~  $and$\\
Department of Physics, Shandong University\\
Jinan, Shandong 250100, P.R.China}

\maketitle

\begin{abstract}
\baselineskip 19pt

 According to the basic properties of QCD, colour reconnection
effects can occur in hadronic processes at high energies.
The comparison of $e^{+}e^{-}\rightarrow
Z^0Z^0\rightarrow q\overline{q}q'\overline{q'}$ events with the superposition
of $Z^0\rightarrow q\overline{q}$ and $Z^0\rightarrow q'\overline{q'}$ events
from LEP1 would provide an unambiguous model-independent probe.
We show  that at LEP2 energy, the background processes are negligible if we 
select only 
$e^{+}e^{-}\rightarrow Z^0Z^0\rightarrow b
\overline{b}c\overline{c}$ events, 
and limit the measurements in the phase space of on-shell $Z^0Z^0$
events.

~

~


keywords: ~~colour-reconnection, $Z^0Z^0$ events, phase space

\end{abstract}

\newpage



The usual treatment of the hadronic processes in high
energy collisions is divided into two phases, the perturbative
parton cascade and the non-perturbative hadronization; 
and a phenomenological colour flow model(CFM) is used to assign the
colour connections of the partons at the end of the first phase.
These colour connections are the 
interface between the two phases, which
 is  the starting point for the second phase.
But it has been realized(see, e.g. \cite{frib} and references therein)
 that CFM
is a good approximation only when $N_C\rightarrow \infty $.   
With $N_{C}=3$ as it is 
in QCD, colour connections of the partons can occur in many different ways.
For example,   
for the $q\overline{q}+ng$($n>1$) system,
as the final states of the
perturbative phase in $e^+e^-$ annihilation,
strict PQCD calculations show\cite{wang1}
that many different parallel complete colour singlet sets
can exist. 
Each of these sets can equivalently act as the bases 
of the colour space of the perturbative  final state,
but none of them is equivalent to the colour 
flow chain obtained from  CFM\cite{wang2}. PQCD cannot 
 tell from which colour singlet set 
the  hadronization starts. 
This implies that colour reconnection effects can be very significant 
in some cases.


A well known example which has been studied frequently in literature is  
the colour reconnection 
effects in $e^{+}e^{-}\rightarrow
W^{+}W^{-}\rightarrow hadrons$ at LEP2 energy. 
Here we have  $e^{+}e^{-}\rightarrow W^{+}W^{-}\rightarrow q_1 
\overline{q}_2q_3\overline{q}_4$, 
and the two initial colour-singlet systems $q_1 
\overline{q}_2$ (from $W^{+}$) and $q_3\overline{q}_4$ (from $W^{-}$) may be
produced closely in space-time. The basic properties of QCD allow the colour
reconnection to occur among partons from $W^{+}$ and $W^{-}$ showers. Such
effects destroy  the naive picture of independent evolution and fragmentation
of $q_1\overline{q}_2$ and $q_3\overline{q}_4$, respectively. 
This is one of the most important sources of the theoretical uncertainty in
determining the W mass and has attracted much attention in recent years
(see \cite{frib,ellis,khoze,si} and references therein). 
But up to now, the manifestations of colour reconnection effects in the final
state hadrons can only
be studied with the help of model-dependent Monte-Carlo events generators 
in which a number of
approximations and/or assumptions are made. Some of 
these approximations and/or assumptions can be
very sensitive to the colour reconnection effects.
So the theoretical uncertainty is usually very large. 
Hence, a very crucial 
and urgent question in this study  is 
how to probe such kind of effects in a model independent manner.

One candidate of such probe currently being discussed\cite{khoze}
 is to compare 
pure hadronic decay $W^{+}W^{-}\rightarrow
q_1\overline{q}_2q_3\overline{q}_4$, in which reconnection 
 between $W^+$ and $W^-$ sources can occur, 
with semileptonic decay 
$W^{+}W^{-}\rightarrow q_1\overline{q}_2l\nu $ of 
the same kinematics. But as pointed out in \cite{khoze},
this comparison  
still suffers from strong dependence on 
the hadronization scenario and on the
choice of model parameters. Moreover, results 
may be strongly sensitive to 
the adopted experimental strategy.
On the other hand, above the $Z^0Z^0$ threshold
at LEP2, the $e^{+}e^{-}\rightarrow Z^0Z^0\rightarrow
q\bar{q}q'\bar{q'} \rightarrow hadrons$ 
events seem to be the most promising candidate
in this aspect\cite{ball, hywj}. Here the hadronic $Z^0Z^0$
events can contain  similar colour reconnection effects
as those in  $W^+W^-$ events,
while the single $Z^0 \rightarrow q\bar{q} \rightarrow hadrons$ 
data can be obtained 
from LEP1 without any theoretical uncertainty and with high precision. 
So the comparison of the 
experimental results in $e^{+}e^{-}\rightarrow Z^0Z^0\rightarrow
hadrons$ and the convolution of two single hadronic $Z^0$ event data from 
LEP1 will allow an unambiguous probe of the colour reconnection effects. 

However, there is a great difficulty here, i.e. 
the background of $Z^0Z^0$ events is too
large. For example, the cross sections of
the corresponding $W^+W^-$ process and 
corresponding QCD process($e^+e^-$ annihilating into 
any $q\bar{q}$ pair) 
are both more than one order of magnitude larger
than that of the signal process.
Other electroweak processes like $e^+e^-$ annihilating through 
$\gamma^* Z^0$ and $\gamma^*\gamma^*$ into four quarks are also 
 significant
\cite{CERN}. Can we reduce the background by selecting some specific type of
events and/or limiting the measurement in some kinematic region?
In  this letter, we show that this question can be answered 
in the affirmative. By selecting the $e^{+}e^{-}\rightarrow Z^0Z^0\rightarrow
b\overline{b}c\overline{c}$ events as the signal process and
limiting the measurements in the phase space of on-shell $Z^0Z^0$ decays,
which is the phase space for most of the $Z^0Z^0$ events above threshold, 
the contribution of background process can indeed be suppressed very much.
Our numerical results show  that they are even negligible
compared to the signal process
$e^{+}e^{-}\rightarrow Z^0Z^0\rightarrow
b\overline{b}c\overline{c}$.
 

For the sake of explicity, we divide the background processes of 
hadronic $Z^0Z^0$ events  
into the following three types:
1.) the corresponding $W^+W^-$ process, 
2.) the corresponding QCD process and
3.) other elctroweak process.
As pointed out above, the total cross section for  them are 
more than one order of magnitude
larger than that of the signal process. 
But these backgrounds can be greatly suppressed
if the measurements are restricted in the following way.

First, $Z^0$ is neutral, while $W^+W^-$ are charged.
If we look only at 
\begin{equation}
 e^+e^- \rightarrow Z^0Z^0 \rightarrow b\bar{b}c\bar{c} \label{zz}
\end{equation}
events as signal 
\footnote{Here we select $b\bar{b}c\bar{c}$ events but not
$b\bar{b}b\bar{b}$ or $c\bar{c}c\bar{c}$ to avoid identical particle
effects. However, if some colour reconnection effects have no relation
with identical particle effeccts, the $b\bar{b}b\bar{b}$ and
$c\bar{c}c\bar{c}$ events, which will not be created via $W^+W^-$ decay
at all, can be included in those measurements. The following qualitative
and quantitative discussions are quite the same, while the events
which can be selected in experiments are increased to about 3 times.}$^,$
\footnote{We also note that heavy quarks are created only in the perturbative 
phase and can be easily  identified in experiments.},
the corresponding 
$W^+W^-\rightarrow c\bar{b}b\bar{c}$ events
are suppressed by the CKM matrix element $|V_{cb}|^2\sim 0.0016$\cite{pdg}.
This leads to  
a strong suppression of five order of magnitude.
So we can neglect $e^+e^-\rightarrow W^+W^- \rightarrow b\bar{b}c\bar{c}$
in comparison with the signal process(1)
(hereafter,we will call the signal process(\ref{zz}) and 3.) other
electroweak processes as 
all the EW process).

Second, comparing the signal process(\ref{zz}) to those in the 
remaining background processes, i.e.

2.) the QCD process 
\begin{equation} \label{qcd}
\begin{array}{lcc}
e^+e^-\rightarrow \gamma^*/Z^0\rightarrow (b\bar{b})/(c\bar{c})&+& g^*\\
& & \downarrow\\ & & (c\bar{c})/(b\bar{b})
\end{array}
\end{equation}

3.) other electroweak processes 
\begin{equation}\label{zg}
e^+e^-\rightarrow \gamma^*Z^0 \rightarrow  b\bar{b}c\bar{c},
\end{equation}
\begin{equation}\label{gg}
e^+e^-\rightarrow \gamma^*\gamma^* \rightarrow  b\bar{b}c\bar{c},
\end{equation}

\begin{equation}\label{oth}
\begin{array}{lcc}
e^+e^-\rightarrow \gamma^*/Z^0\rightarrow
(b\bar{b})/(c\bar{c})&+&\gamma^*/Z^0\\
& &\downarrow \\ & & (c\bar{c})/(b\bar{b}).
\end{array}
\end{equation}
\noindent
the $b\bar{b}$ and $c\bar{c}$ in (\ref{zz}) have the following 
important pecularity: They
 originate predominantely from on-shell $Z^0$ bosons. 
This determines that the 
phase space of each of such kind of quarks and antiquarks is very limited: 
It is easy to see that,
at $\sqrt{S}=2M_Z$(where $M_Z$ is $Z^0$ mass), 
the  velocities of the two on-shell $Z^{0}$'$s$,
$\beta=\sqrt{1-4M_Z^2/S}$, are both zero.
Thus this four quarks and antiquarks 
from on-shell $Z^0$ decay have the same energy 
$M_Z/2$. But for $\beta =0$, 
the on-shell $Z^0Z^0$ cross section $\sigma_{ZZ}=0$.
Therefore we should study the process 
(\ref{zz}) above $Z^0Z^0$ 
threshold, i.e. $\sqrt{S}> 2M_Z$. In this case, 
if the quark mass is neglected,
the quark energy $E_i$($i=c,~\bar{c},~b,~\bar{b}$) in 
the on-shell  $Z^0Z^0$ decay 
satisfies,
\begin{equation}\label{condt}
\frac{M_Z (1-\beta)}{2 \sqrt{1-\beta^2}} \leq E_i
\leq \frac{M_Z (1+\beta)}{2 \sqrt{1-\beta^2}},~~
\end{equation}
This shows that $E_{i}$ is limited to a given small region.
Obviously, this range
becomes wider as $\sqrt{S}$ increases, but even at the highest energy of
LEP2, i.e. $\sqrt{S}=200~GeV$, we have
\begin{equation}\label{cond}
29.5~GeV \leq E_i \leq 70.5~GeV,~~~
\end{equation}
which is still a  very narrow region
compared to the processes (\ref{qcd})-(\ref{oth}), 
in which the quark energy can range from $0$ to $\sqrt{S}/2$, 
since there are non-resonant intermediate states
(eg. $\gamma^*$, $Z^{0*}$, $g^*$).
If 
only the energy range given by  Eq. (\ref{condt}) 
is considered in measurement,
the phase space of the four quarks
in background processes (\ref{qcd})-(\ref{oth}) is restricted into 
a small part of the total, so that their cross sections are all strongly 
suppressed.

To show these effects quantitatively, we
 calculate the cross sections and differential cross sections  of 
processes(1)-(5) in groups.
For the comparison of  the signal process to its background,
(1) should be studied separately.
In the QCD process(\ref{qcd}),
one of the two quark pairs  originates  from a
colour-octet gluon, so both $c\bar{c}$ 
and $b\bar{b}$ are in colour-octets.
In all the other processes, they 
are in colour-singlets. Therefore
there is no interference between process(2) and 
the others. 
Hence the QCD process(2) can be studied separately. 
While all 
the other processes we consider here[(1) and  (3)-(5)] 
can lead to the state with exactly the same quantum numbers, thus
can interfere with each other, so we 
should include 
the contributions 
from all of the interference terms.
We note that the scattering matrices for these processes
can  be calculated using the perturbation theory
in the standard model for electroweak and strong interactions.
The numerical results for the cross sections are obtained 
by integrated in the 
8-dimensional phase space of four fermion system
which is parameterized as usual(see, e.g.\cite{pase}).

Before we present the calculated results  for the cross sections,
we would like to mention the following.
we note that 
in general, 
a  quark  fragments into
a hadron jet which can be observed in experiments. But if
a quark is soft or collinear with other quarks, 
this quark cannot form a resolvable 
hadron jet. There are several different schemes
 to define a resolvable hadron jet. 
We use the Durham scheme\cite{cata}.
According to this scheme, a quark $i$ and another quark $j$ can form 
two resolvable jets if their energies $E_{i}$, $E_{j}$ and 
the angle $\theta_{ij}$ between their moving directions satisfy
the condition $y_{ij}>y_{cut}$. Here   
\begin{equation}
y_{ij}=\frac{2min(E^{2}_{i},E^{2}_{j})(1-cos\theta_{ij})}{S},
         \hspace{0.1in}          (i\neq j)
\end{equation}
and $y_{cut}$ is taken as 0.0015, as an example, same as \cite{ball}. 
If $y_{ij}>y_{cut}$ for all possible 
permutations of $i$ and $j$,
we say that these four quarks fragment 
into four different hadron jets.
This criteria is applied in our calculation, 
thus the numerical results
we present in the following should 
 be understood as those for 
$b\overline{b}c\overline{c}$ four jets.


We now  take $\sqrt{S}=200~GeV$ as an example and
show the results for 
the total cross sections of the $Z^0Z^0$ process(1), 
the corresponding QCD process(2)
and all EW processes[(1) and (3)-(5)] in Fig.1. The shaded areas
represent the corresponding
 cross sections 
$\sigma ^{c}_{ZZ}$, $\sigma ^{c}_{EW}$, $\sigma ^{c}_{QCD}$ 
with the constraint(7) for the phase space.
From the figure, 
we see clearly that the $Z^0Z^0$ events
dominate the $b\overline{b}c\overline{c}$ four jet events,
especially in the phase space as limited in (7).
More precisely, we see that  
the $Z^0Z^0$ events in (7) take about 
94\%~ of the whole(including off-shell) $Z^0Z^0$ events,~ 
this implies our criteria has selected  most of the $Z^0Z^0$ events;
 and that 
$\sigma _{ZZ}^c ~\sim ~\sigma _{EW}^c$, with $\delta _{EW}=\frac{\sigma
_{EW}^c-\sigma _{ZZ}^c}{\sigma _{ZZ}^c}\sim 3\%$, this  shows clearly 
that in the limited phase space  contributions 
from the  non-$Z^0Z^0$ EW processes and all interference
between every two single processes are negligible. 
We also see    
$\delta _{QCD}=\frac{\sigma _{QCD}^c}{\sigma _{ZZ}^c}\sim 0.7\%$. 
This  means that the corresponding QCD process is really negligible.

In Fig. 2 we show the corresponding energy 
distributions $d\sigma /dE$.
In the case that the quark mass are neglected, 
the symmetry of the matrix element  
under the interchange of 
(anti)quark labels implies that the 
energy distributions are identical 
for the four  quarks and antiquarks in the  above mentioned processes.
From this figure, we see again the dominance of $Z^0Z^0$ events in
the energy range(7): 
The distribution curve of $Z^0Z^0$ process and that of all the EW processes 
show a high platform, and they almost coincide in this range.
While outside rang(7), the $Z^0Z^0$ curve drop very fast and the other 
process become dominant.
For the QCD process, the energy 
distribution is peaked near the  two edges: 
$E_j=0$ and $E_j=\sqrt{S}/2$,
since in these events we have quarks 
both from the decay of $\gamma^*/Z^{0*}$ with 
virtuality $\sqrt{S}$(and thus the 
quark energy $E_j$ is $\sqrt{S}/2$) 
and those from soft virtual gluons
(i.e. $E_j\rightarrow 0$). 
but it is an order of magnitude lower than that 
of the $Z^0Z^0$ and the EW processes under the range (\ref{cond}).
These results  show clearly the efficiency of the restriction(\ref{cond}).
It picks most of the signal events
 but drops a large part of the background events.

It may be also interesting to look at the angular distribution
of these processes because of the following. 
In the $Z^0Z^0$ process
(\ref{zz}), the colour 
reconnection may lead to two new colour
singlets $b\bar{c}$ and $c\bar{b}$
if $b$ and $\bar{c}$($c$ and $\bar{b}$) are 
sufficiently close to each other in phase space.
This means that the colour reconnection has large 
probability to occur if
 the angle $\theta$ between $b$ and $\bar{c}$
(or $c$ and $\bar{b}$) is small.
We show in Fig. 3 the distribution
$d\sigma /d\cos\theta$ versus angular 
separation between $c$ and $\bar{b}$
for the $Z^0Z^0$ process, the QCD process and the whole
EW processes under condition (\ref{cond}).
The angular distributions for the $Z^0Z^0$ process 
and the EW processes
almost coincide, the same feature as in Fig 2. 
For the QCD process,
the $c\bar{b}$ angular distribution is 
peaked in the back-to-back
direction, which reflects the dominance 
of the back-to-back configuration for 
$c\bar{b}$(and also $b\bar{c}$) with 
the virtual gluon preferentially
emitted along the quark or antiquark 
directions. For $\cos\theta
\rightarrow 1$, the distribution is 
strongly suppressed by $y_{cut}$,
while for $\cos\theta \rightarrow -1$, 
it is slightly restricted by
the condition (\ref{cond}). Obviously, 
when $\theta$ is less than about 53 degree,
the differential cross section of the QCD process is 
at least three orders of magnitude lower than the $Z^0Z^0$
one, which makes the measurements of
the possible colour reconnection effects in 
the real $Z^0Z^0$ process more feasible.

In summary, the study of the colour reconnection effects
in hadronic reactions is of fundamental significance in 
understanding QCD. 
High statistical data above the $Z^0Z^0$ threshold
at LEP2 may allow   a model-independent probe of these effects. 
The background processes are suppressed to a negligible
level if we choose only $e^+e^- \rightarrow Z^0Z^0 \rightarrow
b\overline{b}c\overline{c}$ as signal events and limit the measurement 
in the energy range given by (6).
Qualitative analysis and quantitative results
presented in this letter show the following:
 First, the greatest 
pollution to the signal process, the corresponding $W^+W^-$
process, can be dropped by the CKM suppression for $c,\bar{b}$
($b,\bar{c}$) pair.
Second, in energy range(\ref{condt}),
the pollution from the corresponding QCD process,  
that from other electroweak  processes and that from  all 
electroweak  interference terms are  
negligibly small, while most of the $Z^0Z^0$ events are picked.
Furthermore, limiting the angle between 
$b$ and $\overline{c}$(or between $c$ and $\overline{b}$) 
to small values,
where colour reconnections occur with larger possibilities,
the QCD background will be further suppressed.
So comparing $e^{+}e^{-}\rightarrow Z^0Z^0\rightarrow b \overline{b}c
\overline{c}$ events with the superposition of
the corresponding $Z^0\rightarrow b\overline{b}$
and $Z^0\rightarrow c\overline{c}$ events from LEPI would provide
an unambiguous model-independent probe of the colour reconnection effects.

ACKNOWLEDGEMENT

We thank P. de Jong, Z. Liang, W. Metzger
 and  T. Sj\"{o}strand for helpful  discussions.
This work is supported in part by National Natural Science Foundation
of China(NSFC).  

\newpage


\newpage

FIGURE CAPTIONS

Fig. {\bf 1}. The total cross section for  $e^+e^- \rightarrow
Z^0 Z^0 \rightarrow b\bar{b}c\bar{c}$, that for                                     
the whole EW processes and that for the corresponding QCD process
at $\sqrt{S}=200GeV$. The shaded area represents the corresponding
cross section under the restriction $29.5GeV<E_{i}<70.5GeV$.
In  the calculations here and following $\alpha_{s}$ is set to 0.1.

Fig. {\bf 2}.
The energy distribution $\frac{d\sigma}{dE_b}$ of
the $Z^0 Z^0$ process(solid line),
the whole EW processes(dashed line) and the QCD process(dotted line).
The dash-dotted line denotes the edges of the
energy range $29.5GeV<E<70.5GeV$.

Fig. {\bf 3}.
The angular distribution $\frac{d\sigma}{d\cos\theta}$
of the $Z^0 Z^0$ process(solid line),
the whole EW processes(dotted line) and the QCD process(dashed line)
under the restriction $29.5GeV<E<70.5GeV$.

\newpage

\begin{figure}
\psfig{file=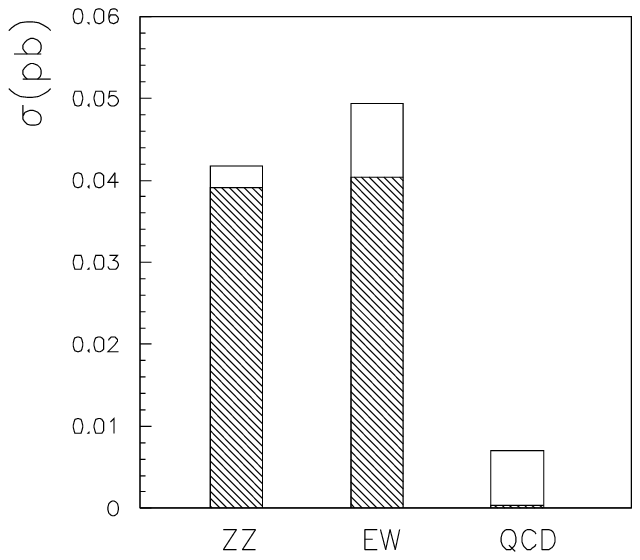,height=10cm,width=10cm}
\caption{~}
\end{figure}

\newpage
\begin{figure}
\psfig{file=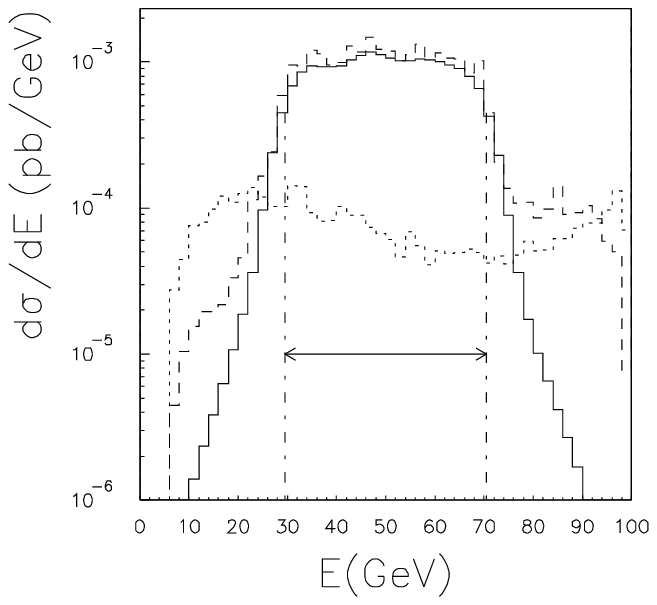,height=10cm,width=10cm}
\caption{~}
\end{figure}

\newpage
\begin{figure}
\psfig{file=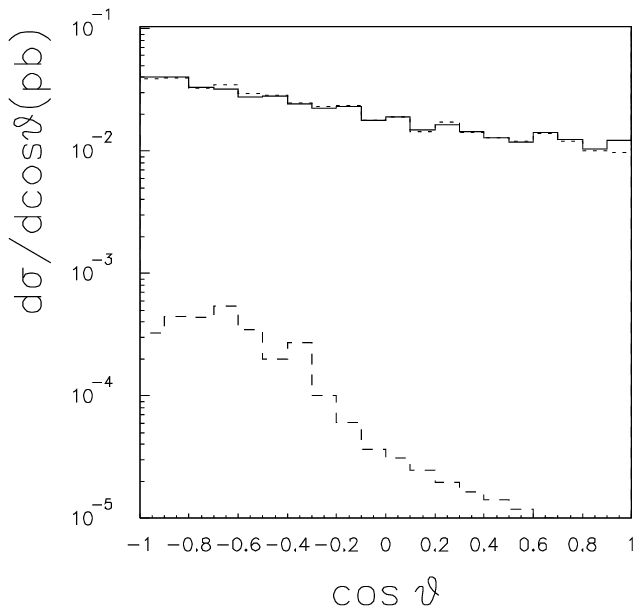,height=10cm,width=10cm}
\caption{~}
\end{figure}

\end{document}